# Analyzing the Impact of Financial Inclusion on Economic Growth in Bangladesh.

**Ganapati Kumar Biswas, Senior Principal Officer & Company Secretary, Pubali Bank Securities Limited, Bangladesh. E-mail: ganapatibiswas@gmail.com.**

**Abstract**: Financial inclusion is touted one of the principal drivers for economic growth for an economy. The study aims to explore the impact of financial inclusion on economic growth in Bangladesh. In my study, I used the number of loan accounts as the proxy for financial inclusion. Using time series data from spans from 2004-2021, the study revealed that there exists a long-run relationship between GDP, financial inclusion, and other macroeconomic variables in Bangladesh. The study also found that financial inclusion had a positive impact on economic growth of Bangladesh during the study period. Therefore, the policymakers and the central bank of Bangladesh as the apex authority of financial system should promote financial inclusion activities to achieve sustainable economic growth.

1. Introduction

Financial inclusion has been identified as one major forces for achieving sustainable economic growth by academicians and researchers across globe (Ali et al., 2020; G. K. Biswas & Ahamed, 2023; Faruq, 2023; Lenka & Sharma, 2017). In a theoretical perspective, it has been argued that financial inclusion is a drive forces toward economic growth. The earlier research has demonstrated that financial inclusion boosts economic growth. Financial sector via its services not only help for the accessibility of capital formation but also encourage innovation, efficiency, and investment which in turn into growth in output. For achieving sustainable growth, the banking sector must play an important role, and the banking and financial services must be cost effective since it encourages capital accumulation and attract business competition among the banks which result the more investment and growth.

The concept of financial inclusion is comparatively new. In the early 2000s, the concept of financial inclusion was introduced, which was derived from research that highlighted the direct consequences of financial exclusion on poverty and low economic growth. The motivation for financial inclusion is to ensure that all adults in society have easy access to large-scale financial products that are personalized to their needs and provided at reasonable cost. These products include payments, savings, credit, insurance, and pensions. The central bank of any country as the regulator authority of financial system has a big role to play for building the financially inclusive society (Anjom & Faruq, 2023; G. Biswas, 2023; King & Levine, 1993).

Many studies have been conducted to explore the impact of financial inclusion on economic growth in different countries' perspective. The objective of this paper is to assess the impact of financial inclusion on economic growth in Bangladesh.

## 2. Literature Review

A considerable number of studies have tried to explore the effect of financial inclusion on economic growth (Ahmad et al., 2023; Singh & Ghosh, 2021; Van et al., 2021).

Pearce (2011) analyzed the impact of financial inclusion on growth in MENA region. The study confirmed that for increasing income, decreasing poverty, and creating employment. The study also mentioned that the access to financial services is in MENA region is limited and their financial inclusion is characterized by NGO-dominated microcredit sectors. The study also states that improvements in financial literacy is still limited to few countries.

Kim et al. (2018) examined the relationship between financial inclusion and economic growth in Organization of Islamic Cooperation (OIC) countries. By using the panel data for 55 OIC countries the study found that that financial inclusion had a positive impact on economic growth in these countries. The study also revealed that financial inclusion and economic growth have mutual causalities with each other. Finally, the study concluded that financial inclusion had positive effect on the economic growth in OIC countries.

Biswas (2023) examined the contribution of financial inclusion on economic growth in 4 South Asian countries. The study used panel data models, and several measures of financial inclusion to explore the relationship between economic growth and financial inclusion. The results of the study confirmed that financial inclusion had a positive impact on economic growth in those countries although this effect varies across different measures of financial inclusion. The study also suggested that policy makers in these countries must take necessary steps to accelerate financial inclusion activities to achieve robust economic growth.

Dixit & Ghosh (2013) conducted a study on financial inclusion on different states in India. The study concluded that financial inclusion strategy should be implemented uniformly across states in the country to attain comprehensive growth. The study also mentioned that financial inclusion can be a means for reducing poverty and income inequality.

## 3. Data and Methodology

The aim of this paper is to explore the role of financial inclusion on economic growth in Bangladesh. For conducting the study, I have collected the data of financial inclusion from Financial Access Survey (FAS) of the International Monetary Fund. Other data related to this study are derived from the World Development Index (WDI) of the World Bank.

For this study, I have estimated the following equation:

$$lgdp_t = \alpha_0 + \alpha_1\, lac_t + \alpha_2\, lfdi_t + \alpha_2\, lhc_t + \varepsilon_t \qquad (1)$$

where, $lgdp_t$ denotes logarithm of Gross Domestic Product (GDP), $lac_t$ is the logarithm of number loan accounts with commercial banks, which is the measure of financial inclusion. $lfdi_t$ and $lhc_t$ refer to natural logarithm of Foreign Direct Investment (FDI) and natural logarithm of human capital respectively, and finally $\varepsilon_t$ is the error term.

## 4. Estimated Results
### 4.1. Trend of Variables

Time series data suffers from the risk of unit root problem and this problem might arise if the series have clear trend. Therefore, before conducting, formal unit root test I check whether the series have clear trend or not. The below Figure 1 shows the graph of log of GDP (l_lgdp) and log of loan account (l_lac) of Bangladesh.

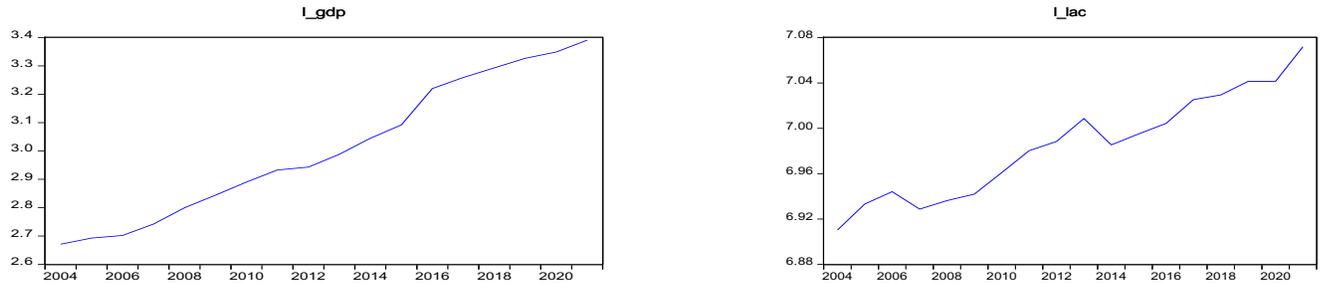

**Figure 1: Graph of l_gdp and l_lac**

From figure 1, it is evident that both variables have clear trend, so the series might have unit root problem i.e. the series might be non-stationary. Therefore, in the following section, I conducted formal unit root test.

### 4.2. ADF Test

In this section, I have conducted the Augmented Dickey-Fuller (ADF) to check whether series have unit root problem or not. The Table 1 below shows the result of ADF test.

| Variable | Parameter | t-statistics | P-Value | Decision |
|---|---|---|---|---|
| L_lac | Level | -2.8094 | 0.2128 | I(1) |
|  | First diff. | -4.4423 | 0.0148 |  |
| l_gdp | Level | -1.2974 | 0.8359 | I(1) |
|  | First diff. | -5.5127 | 0.0050 |  |
| l_fdi | Level | -2.0091 | 0.5555 | I(1) |
|  | First diff. | --5.9894 | 0.0011 |  |
| l_hc | Level | -2.5236 | 0.3137 | I(1) |
|  | First diff. | -4.3548 | 0.0172 |  |

**Table 1: Results of ADF Test**

From Table 1, we couldn't reject the null hypothesis that the series has unit root at level for all variables as the p-values are greater than 0.10. But at first difference, we reject the null hypothesis at 5% significance level as the p-values are less than 0.5. So, we conclude that all the variables are I(1). Therefore, in the following section, I conduct the Johansen cointegration test to check whether the series are integrated or not.

### 4.3. Johansen Cointegration Test

The below Table 2 presents the results of Johansen cointegration test. This test has been carried out to check the cointegration among variables.

Unrestricted Cointegration Rank Test (Trace)

| Hypothesized No. of CE(s) | Eigenvalue | Trace Statistic | 0.05 Critical Value | Prob.** |
|---|---|---|---|---|
| None * | 0.944838 | 73.48422 | 47.85613 | 0.0000 |
| At most 1 | 0.672883 | 27.12461 | 29.79707 | 0.0986 |
| At most 2 | 0.391638 | 9.245623 | 15.49471 | 0.3432 |
| At most 3 | 0.077682 | 1.293850 | 3.841465 | 0.2553 |

Trace test indicates 1 cointegrating eqn(s) at the 0.05 level

Unrestricted Cointegration Rank Test (Maximum Eigenvalue)

| Hypothesized No. of CE(s) | Eigenvalue | Max-Eigen Statistic | 0.05 Critical Value | Prob.** |
|---|---|---|---|---|
| None * | 0.944838 | 46.35961 | 27.58434 | 0.0001 |
| At most 1 | 0.672883 | 17.87899 | 21.13162 | 0.1344 |
| At most 2 | 0.391638 | 7.951773 | 14.26460 | 0.3836 |
| At most 3 | 0.077682 | 1.293850 | 3.841465 | 0.2553 |

Max-eigenvalue test indicates 1 cointegrating eqn(s) at the 0.05 level

**Table 2: Johansen Cointegration Test**

From above Table 2, we see that both Trace test and Max- Eigen Test indicates that all the variables are cointegrated i.e. these variables move together in the long-run.

### 4.4. Vector Error Correction Model (VECM) Estimations

From previous section, we saw that all the variables are cointegrated. Hence, I have estimated the VECM and the results of VECM is shown below:

| Error Correction: | D(L_GDP) | D(L_LAC) | D(L_FDI) | D(L_HC) |
|---|---|---|---|---|
| D(L_GDP(-1)) | -0.075169 | 0.071253 | -2.313933 | -0.102445 |
| | (0.27899) | (0.15245) | (1.45102) | (0.11565) |
| | [-2.26944] | [ 0.46739] | [-1.59469] | [-0.88582] |
| D(L_LAC(-1)) | -0.519902 | -0.164922 | -4.332142 | -0.152170 |
| | (0.56980) | (0.31136) | (2.96352) | (0.23620) |
| | [-2.31244] | [-0.52969] | [-1.46182] | [-0.64424] |
| D(L_FDI(-1)) | 0.015320 | 0.020239 | -0.318263 | -0.066568 |
| | (0.05973) | (0.03264) | (0.31065) | (0.02476) |
| | [ 0.25650] | [ 0.62010] | [-1.02452] | [-2.68862] |
| D(L_HC(-1)) | 0.719094 | 0.090676 | 1.827206 | -0.311230 |
| | (0.46018) | (0.25146) | (2.39340) | (0.19076) |
| | [ 1.56264] | [ 0.36060] | [ 0.76343] | [-1.63152] |
| C | 0.042152 | 0.005270 | 0.143109 | 0.024317 |
| | (0.01464) | (0.00800) | (0.07612) | (0.00607) |
| | [ 2.88006] | [ 0.65890] | [ 1.88002] | [ 4.00799] |
| R-squared | 0.291200 | 0.177918 | 0.408050 | 0.680465 |
| Adj. R-squared | -0.063200 | -0.233123 | 0.112075 | 0.520698 |
| F-statistic | 0.821670 | 0.432848 | 1.378662 | 4.259097 |

Table 3: VECM Estimations

The Table 3 above shows the results of the VECM estimates. Each column shows the equation for each endogenous variable in the model. For example, the only statistically significant determinant of the human capital is the lagged value of the foreign direct investment [l_fdi (–1)]. This means that this year's human capital can be duly estimated by our knowledge of the foreign direct investment in the previous year. On the other hand, the log of GDP is determined lagged value of itself and money supply and one period the lagged value of number of loan account.

5. Conclusion

The objective of the paper was to examine the impact of financial inclusion measured by the number of loan accounts on economic growth in Bangladesh. Using time series data from 2004-2021, the study revealed that there exists a long-run relationship between GDP, number of loan account and other variables. The study also found that the financial inclusion had a positive impact on economic growth of Bangladesh during the study period. Therefore, the policymakers and the central bank of Bangladesh as the apex authority of Bangladesh should promote financial inclusion activities to achieve sustainable economic growth.